\documentclass[prd,preprint,superscriptaddress,showpacs,byrevtex]{revtex4}
\usepackage{epsfig}

\def\as{\alpha_s}
\def\gl{\tilde{g}}
\def\sq{\tilde{q}}
\def\sqb{\bar{\tilde{q}}}
\def\qb{\bar{q}}

\def\ms{m_{\tilde q}}
\def\mg{m_{\tilde g}}

\def\md{m_{-}}

\def\ghat{\hat{g}_s}

\def\bs{\beta_{\sq}}

\def\bg{\beta_{\gl}}

\newcommand{\Lg}[2]{\log\left(\frac{#1}{#2}\right)}

\newcommand{\be}{\begin{equation}}
\newcommand{\ee}{\end{equation}}
\newcommand{\bea}{\begin{eqnarray}}
\newcommand{\eea}{\end{eqnarray}}
\def\simgt{\rlap{\lower 3.5 pt \hbox{$\mathchar \sim$}} \raise 1pt
Ê \hbox {$>$}}

\def\as{\alpha_s}

\def\ms{m_{\tilde q}}
\def\mg{m_{\tilde g}}
\def\md{m_{-}}

\begin{document}

\title{String Theory at LHC Using Supersymmetry Production From String Balls }

\author{Gouranga C. Nayak}\email{nayak@max2.physics.sunysb.edu}

\affiliation{ 22 West Fourth Street \#1, Lewistown, Pennsylvania 17044, USA }
\date{\today}

\begin{abstract}

If extra dimensions are found in the second run of LHC in the $pp$ collisions at
$\sqrt{s}$ = 14 TeV then the string scale can be $\sim$ TeV, and we should produce
string balls at LHC. In this paper we study supersymmetry (squark and gluino)
production from string balls at LHC in $pp$ collisions at $\sqrt{s}$ = 14 TeV
and compare that with the parton fusion results using pQCD. We find
significant squark and gluino production from string balls at LHC which is
comparable to parton fusion pQCD results. Hence, in the absence of black
hole production at LHC, an enhancement in supersymmetry production
can be a signature of TeV scale string physics at LHC.

\end{abstract}
\maketitle

\newpage

\section{Introduction}

The large hadron collider (LHC) has completed its first run where we have not found any
evidence of the beyond standard model physics in $pp$ collisions at $\sqrt{s}$ = 7 TeV and
8 TeV respectively \cite{atlas}. The LHC has started its second run involving $pp$ collisions at $\sqrt{s}$
= 13 TeV and will reach its maximum energy $\sqrt{s}$ = 14 TeV in $pp$ collisions in future. Hence all
our calculations in this paper is performed at the maximum energy in the $pp$ collisions at the LHC,
{\it i. e.}, in $pp$ collisions at
$\sqrt{s}$ = 14 TeV. The LHC will also collide two lead nuclei at $\sqrt{s}$ = 5.5 TeV per nucleon
to produce quark-gluon plasma \cite{qgp,qgp1}. Since the total energy in the two lead nuclei collisions
at LHC will be $\sim$1150 TeV, we may expect to see new physics \cite{newph} in the nuclear collisions
at LHC as well.

If we find extra dimensions in the second run at LHC in the $pp$ collisions at $\sqrt{s}$ = 14 TeV
then the scale of quantum gravity {\it could be} as low as $\sim$
TeV \cite{folks,ppbf,pp,pp1,pp2,pp3,ag,ppch,ppk,ppu,park,hof,more,gramm,cham,pp4,bv,pp6}. In the
presence of extra dimensions, the string mass scale $M_s$ and the
Planck mass $M_P$ could be around $\sim$ TeV. In this situation we can
look forward to search for TeV scale string physics at
CERN LHC.  Also, it is unknown if supersymmetry is manifested in nature, and if SUSY does exist
it is unclear at what energy scale SUSY becomes manifest. Note that, as mentioned above,
we have not found any experimental evidence for string ball production and supersymmetry
production in the first run at LHC \cite{atlas}.
Hence, we can only search for string balls, extra dimensions and superpartners in the second run
at the LHC. If we manage to detect {\it any} of these exotic phenomena, then we will be propelled into the twenty-first
century, as our understanding of quantum gravity and perhaps even string theory is revolutionized.

One of the most exciting possibility is to search for TeV scale black hole
and string ball production at LHC. These `brane-world' black holes and string balls will
be our first window into the extra dimensions of space predicted by string theory,
and required by the several brane-world scenarios \cite{large}.
There may be many other ways of testing string theory at LHC
starting from brane excitations to various string excitations. The
string balls of \cite{dimo} is just one such model, where the predictions
are done in a toy string theory model. In this paper we will focus on studying
string theory at LHC based on string balls.

There has been arguments that the black hole stops radiating near
Planck scale and forms a black hole remnant \cite{dasall}. These black hole
remnants can be a source of dark matter \cite{steinhardt, darknayak}. In the
absence of a theory of quantum gravity, we can study other scenarios
of black hole emissions near the Planck scale. Ultimately, experimental
data will determine which scenarios are valid near the Planck scale.
In this paper we will study string ball production at LHC in the context
of black hole evaporation in string theory. Recently, string theory has given
convincing microscopic calculation for black hole evaporation
\cite{susskind,allstring}.

String theory predicts that a black hole has formed at several times
the Planck scale and any thing smaller will dissolve into some thing
known as string ball \cite{dimo}. A string ball is a highly
excited long string which emits massless (and massive) particles at Hagedorn
temperature with thermal spectrum \cite{amati,canada}. For general relativistic
description of the back hole to be valid, the black hole mass $M_{BH}$ has to be
larger than the Planck mass $M_P$. In string theory the string ball mass $M_{SB}$
is larger than the string mass scale $M_s$. Typically
\bea
&& M_s < M_P < \frac{M_s}{g_s^2}  \nonumber \\
&& M_S << M_{SB} << \frac{M_s}{g_s^2} \nonumber \\
&& \frac{M_s}{g_s^2} << M_{BH}
\label{scale}
\eea
where $g_s$ is the string coupling which can be less than 1
for the string perturbation theory to be valid. Since string ball
is lighter than black hole, more string balls are expected to be
produced at CERN LHC than black holes.

The Hagedorn temperature of a string ball is given by
\bea
T_{SB}=\frac{M_s}{\sqrt{8} \pi}
\label{tsb}
\eea
where $M_s ~\sim $ TeV is the string scale. Since this temperature is very high at LHC
($\sim$ hundreds of GeV) we expect more massive particles ($M \sim 3 T_{SB}$) to be
produced at CERN LHC from string balls.

In this paper we study squark and gluino production from string balls at CERN LHC and
make a comparison with the squark and gluino production from the parton fusion processes
using pQCD. We present the results for the total cross section, rapidity distribution and
$\frac{d\sigma}{dp_T}$ of squark and gluino production. There can be significant squark
and gluino production
from black holes at LHC as well \cite{nayaktop}. This is because the black hole
temperature increases as its mass decreases whereas the string ball temperature
remains constant (see eq. (\ref{tsb})). On the other hand the string ball mass
is smaller than the black hole mass and more string balls are produced at LHC.
Hence squark and gluino production at LHC is from two competitive effects:
1) string ball (black hole) production at LHC and
2) squark and gluino emission from a single string ball (black hole) at LHC.
We find significant squark and gluino production from string balls at LHC. Hence,
in the absence of black hole production at LHC, an enhancement in the supersymmetry
production may be a signature of TeV scale string physics at LHC.

The paper is organized as follows. In section II we discuss string balls
in string theory. In section III we describe
squark and gluino production in pQCD. Section IV describes
squark and gluino production from string balls at the CERN LHC.
We present our results and discussions in section V and conclude in section VI.

\section{ String Balls in String Theory }

In string theory the fundamental scales are as follows: $l_P$ is the Planck length
scale, $l_s$ is the quantum length scale of the string, $\alpha'=l_s^2$ is the
inverse of the classical string tension, $M_s=\frac{1}{l_s}$ is the string mass
scale and $g_s$ is the string coupling. In small string coupling
\bea
l_P \sim g_s l_s.
\label{length}
\eea
For $d=3+n$ space dimensions one obtains
\bea
l_P^{d-1} \sim g_s^2 l_s^{d-1}.
\label{dlength}
\eea

As black hole shrinks it reaches the correspondence point \cite{susskind,allstring}
\bea
M \le M_c \sim \frac{M_s}{g^2_s}
\eea
and makes a transition to a configuration dominated by a highly excited long string
(known as string ball) which continues to lose mass by evaporation
at the Hagedorn temperature \cite{amati} and "puffs-up" to a larger "random-walk" size which
has observational consequences. Evaporation, still at the Hagedorn temperature,
then gradually brings the size of the string ball down towards $l_s$.

At LHC we are interested in production and decay of highly excited string.
Production of a highly excited string from the collision of two
light string states at high $\sqrt{s}$ can be obtained from the Virasoro-Shapiro
four point amplitude
\bea
A(s,t)=\frac{2\pi g_s^2 \Gamma[-1-\alpha's/4]\Gamma[-1-\alpha't/4]\Gamma[-1-\alpha'u/4]}{\Gamma[2+\alpha's/4]
\Gamma[2+\alpha't/4]\Gamma[2+\alpha'u/4]}
\eea
by using string perturbation theory with
\bea
s+t+u =   -16 /\alpha'.
\eea
The production cross section is therefore given by
\bea
\sigma \sim \frac{\pi {\rm Res} A(\alpha's/4 = N, t=0)}{s} = g_s^2 \frac{\pi^2}{8} \alpha'^2 s.
\label{cs1}
\eea

The cross section in eq. (\ref{cs1}) saturates the
unitarity bounds at around $g_s^2 \alpha' s \sim 1$ which implies that the production cross section for
string balls grows with $s$ as in eq. (\ref{cs1}) only for
\bea
M_s << \sqrt{s} << M_s/g_s,
\eea
while for
\bea
\sqrt{s} >> M_s/g_s,~~~~~~~~~~~~~~~~~~~~~~~~~~\sigma_{SB} = \frac{1}{M_s^2}
\eea
which is constant.

The string ball production cross section in a parton-parton collision is therefore given by \cite{dimo}
\bea
&&~ {\hat \sigma}_{SB} \sim \frac{g_s^2 M_{SB}^2}{M_s^4},~~~~~~~~~~~M_s<<M_{SB}<<M_s/g_s, \nonumber \\
&&~ {\hat \sigma}_{SB} \sim \frac{1}{M_s^2},~~~~~~~~~~~M_s/g_s<<M_{SB}<<M_s/g^2_s.
\label{sbc}
\eea

Highly excited long strings emit massless (and massive) particles
at Hagedorn temperature as given by eq. (\ref{tsb}) by using the
conventional description of evaporation in terms of black body radiation \cite{amati}
where the emission can take place either in the bulk (in to
the closed string) or in the brane (in to open strings).

\section{Squark and Gluino Production in pp Collisions at LHC using pQCD }

In this section we discuss the supersymmetry (squark and gluino) production
in quark and gluon fusion processes using pQCD methods applied to high
energy hadronic collisions. The leading order (LO) Feynman diagrams are
given in Fig. 1.

\begin{figure}[htb]
\vspace{2pt}
\centering{{\epsfig{figure=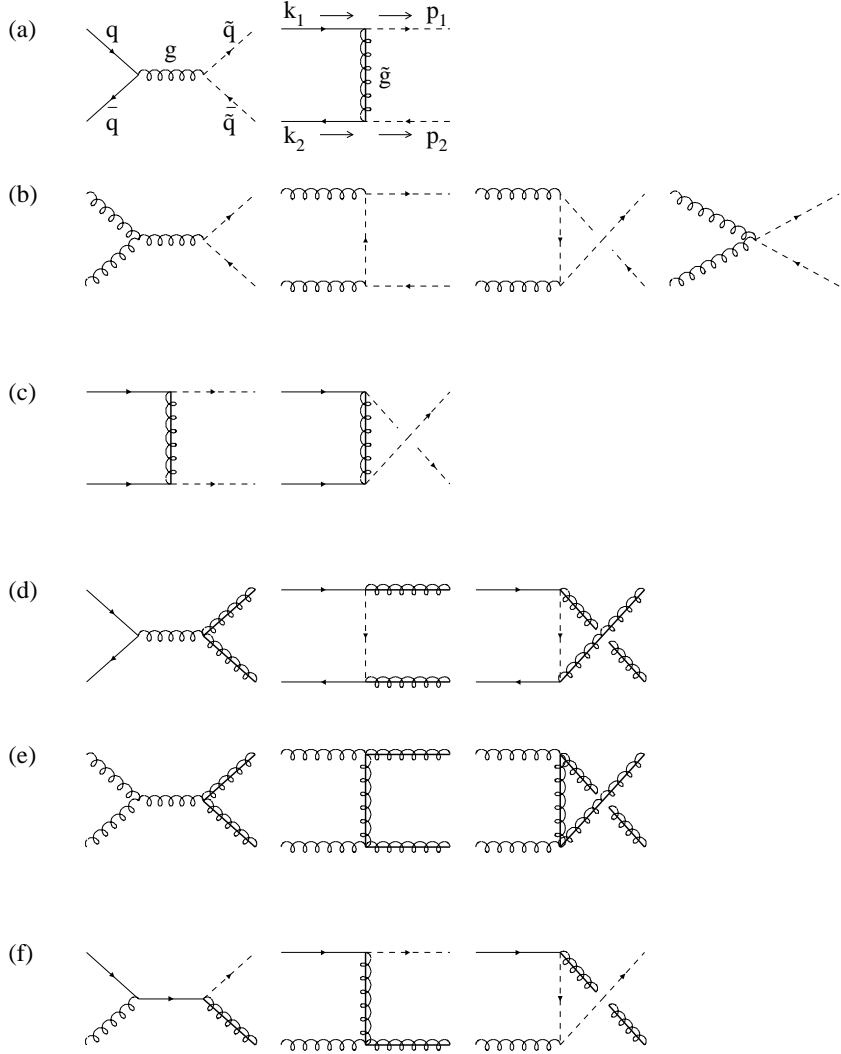,height=17cm}}}
\caption{ Feynman diagrams for the production of squarks and
gluinos in lowest order.}
\label{fig1}
\end{figure}

The differential cross section in the lowest order (LO) is given by:
\bea
\frac{d\sigma}{dp_t^2 dy}~=~\frac{H}{s}~ &&
\sum_{i,j=q,\bar q, g} ~\int_{x_1^{min}}^1 dx_1 ~(- \frac{1}{x_1^2t})
~f_{i/A}(x_1,p_t^2)~f_{j/B}(-\frac{x_1 t s}{x_1s+u},p_t^2)  \times \nonumber \\
&&~\sum
|M|^2(\frac{-x_1^2ts}{x_1s+u},x_1t,-\frac{x_1tu}{x_1s+u})
\eea
where $s$ is the total center of mass energy at hadronic level,
$t=-\sqrt{s(p_t^2+m^2)}~e^y$ and $u=-\sqrt{s(p_t^2+m^2)}~e^{-y}$ where the
lower limit of the $x_1$ integration is given by $x_1^{min}=-u/s+t$.
In the above equation $f_{i/p}(x,Q^2)$ is the parton distribution function inside a proton with
longitudinal momentum fraction $x$ and factorization scale $Q$,
$H=K_{ij}/16\pi$ with $K_{qq}=K_{q\bar q}=1/36$, $K_{gg}=1/256$ and
$K_{qg}=1/96$.

The matrix element squares from partonic fusion processes
at the Born level (see Fig-1) are given by \cite{been}:

\begin{small}
\begin{eqnarray*}
  \sum |{\cal M}^B|^2 (q_i\qb_j\to\sq\sqb) & = &
  \delta_{ij}\left[32 n_f g_s^4  \frac{\hat t_q \hat u_q -\ms^2 \hat s}{\hat s^2}
  + 16 \ghat^4  \frac{\hat t_q \hat u_q -(\ms^2 -\mg^2) \hat s}{\hat t_g^2}
\right. \label{born1} \\
  & &\hphantom{\delta_{ij}a} \left. {} \hspace*{-1cm}
  - 32/3 g_s^2 \ghat^2 \frac{\hat t_q \hat u_q -\ms^2 \hat s}{\hat s \hat t_g}
\right]
  + (1-\delta_{ij})
  \left[ 16 \ghat^4 \frac{\hat t_q \hat u_q -(\ms^2 -\mg^2)\hat s}{\hat t_g^2}\right]
    \nonumber\\[0.2cm]
    \sum |{\cal M}^B|^2 (gg\to\sq\sqb) & = & \!\!4 n_f g_s^4
    \left[24\!\left(1 -2\,\frac{\hat t_q \hat u_q}{\hat s^2}\right) -\! 8/3\right]\! \left[ 1
      -\!2\,\frac{\hat s\ms^2}{\hat t_q \hat u_q}\left(1 -\!\frac{\hat s\ms^2}{\hat t_q
        \hat u_q}\right)\right] \quad\qquad\\[0.2cm]
  \label{squarkborn}
  \sum |{\cal M}^B|^2 (q_i q_j\to\sq\sq) & = &
  \delta_{ij}\Bigg[ 8\ghat^4 \left(\hat t_q \hat u_q -\ms^2 \hat s\right)
  \left(\frac{1}{\hat t_g^2} +\frac{1}{\hat u_g^2}\right)  \\
  & &{} \hphantom{\delta_{ij}a} + 16\ghat^4 \,\mg^2  \hat s \left(  \left(\frac{1}{\hat t_g^2} +\frac{1}{\hat u_g^2}\right)
  - 8/3  \,\frac{1}{\hat t_g \hat u_g}\right) \Bigg]\nonumber\\
  & &{} + (1-\delta_{ij})
  \left[ 16 \ghat^4 \ \,\frac{\hat t_q \hat u_q -(\ms^2-\mg^2) \hat s}{\hat t_g^2}\right]
    \nonumber
\end{eqnarray*}
\begin{eqnarray*}
  \sum |{\cal M}^B|^2 (q\qb\to\gl\gl) & = &
  96 g_s^4 \, \left[\frac{2 \mg^2 \hat s +\hat t_g^2 +\hat u_g^2}{\hat s^2} \right]\\
  & &{} + 96 g_s^2 \ghat^2 \, \left[\frac{\mg^2 \hat s +\hat t_g^2}{\hat s \hat t_q}
  +\frac{\mg^2 \hat s +\hat u_g^2}{\hat s \hat u_q} \right ]\nonumber\\
  & &{} + 2\ghat^4 \left[ 24 \left(\frac{\hat t_g^2}{\hat t_q^2} +
  \frac{\hat u_g^2}{\hat u_q^2}\right) +8/3 \left(2\,\frac{\mg^2 \hat s}{\hat t_q\,\hat u_q} -
  \frac{\hat t_g^2}{\hat t_q^2} - \frac{\hat u_g^2}{\hat u_q^2} \right)\right]\nonumber\\[0.2cm]
  \sum |{\cal M}^B|^2 (gg\to\gl\gl) & = & 576g_s^4   \left(1
  -\frac{\hat t_g \hat u_g}{\hat s^2}\right) \left[\frac{\hat s^2}{\hat t_g\,\hat u_g} -2
  +4\,\frac{\mg^2 \hat s}{\hat t_g \hat u_g} \left (1-\frac{\mg^2 \hat s}{\hat t_g \hat u_g}\right)
  \right]\\[0.2cm]
  \sum |{\cal M}^B|^2 (qg\to\sq\gl) & = &
  2  g_s^2 \ghat^2 \left[ 24 \left( 1 -2\,\frac{\hat s \hat u_q}{\hat t_g^2}
  \right) - 8/3 \right]
  \Bigg[-\frac{\hat t_g}{\hat s} \label{born6} \\
  & &{} + \frac{2(\mg^2-\ms^2)\,\hat t_g}{\hat s \hat u_q} \left(1 +\frac{\ms^2}{\hat u_q}
  +\frac{\mg^2}{\hat t_g} \right) \Bigg], \nonumber
\end{eqnarray*}
\end{small}
where the variables $\hat t_{q,g}$ ($\hat u_{q,g}$) are related to
the Mandelstam variables $\hat t$ ($\hat u$) by:
$\hat t_q~(\hat u_q)$~=~$\hat t~(\hat u)~-~m_{\tilde q}^2$
and
$\hat t_g~(\hat u_g)$~=~$\hat t~(\hat u)~-~m_{\tilde g}^2$ where
$m_{\tilde q}$ ($m_{\tilde g}$) is the mass of the squark (gluino). The $g_s$ is the QCD
gauge coupling ($qqg$) and $\hat g_s$ is the Yukawa coupling ($q {\tilde q}{\tilde g}$),
$n_f$ is the number of flavors and
\be
\hat s ~=~-\frac{x_1^2ts}{x_1s+u}~~~~~~~\hat t_{g(q)}~=~x_1t~~~~~~~~\hat u_{g(q)}~=~-\frac{x_1tu}{x_1s+u}.
\ee

The $p_t$ differential cross sections for squarks and gluinos at LHC are determined
from the above formula by integrating over the rapidity:
\be
\frac{d\sigma}{dp_t}~=~2 p_t~\int_{-cosh^{-1}(\frac{\sqrt s}{2(\sqrt{p_t^2
+m^2}})}^{cosh^{-1}(\frac{\sqrt s}{2(\sqrt{p_t^2+m^2}})}
~dy~\frac{d\sigma}{dm_t^2dy}
\ee
and the rapidity differential cross sections for squarks and gluinos produced at LHC are
given by by integrating over $p_t$:
\be
\frac{d\sigma}{dy}~=~ \int_{m^2}^{s/4cosh^2y}~dm_t^2~\frac{d\sigma}{dm_t^2dy}.
\ee

\begin{figure}[htb]
\vspace{2pt}
\centering{{\epsfig{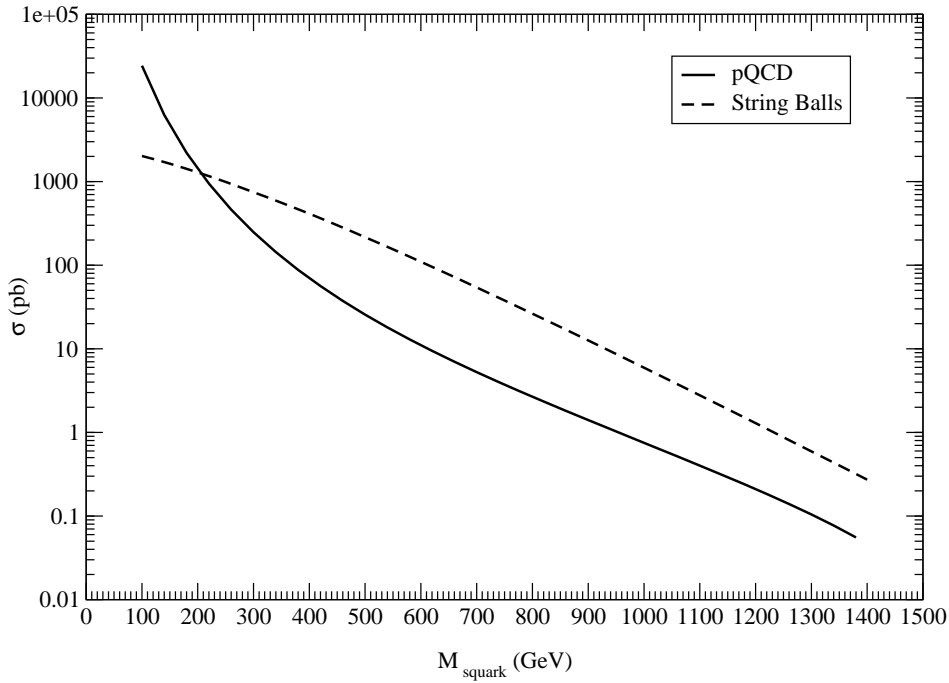}}}
\caption{ Total cross section for squark production at LHC as a
function of squark mass.}
\label{fig2}
\end{figure}

Similarly the total cross sections for squarks and gluinos produced at LHC are then given by:
\begin{equation}
\sigma^{pp \rightarrow {\tilde q} {\tilde {\bar q}} ({\tilde g}{\tilde g})} =
\Sigma_{i,j} \int_{4M^2/s} dx_1 \int_{4m^2/sx_1} dx_2 f_{i/p}(x_1,Q^2) f_{j/p}(x_2,Q^2)
\sigma^{ij}(\hat s)
\end{equation}
where $\sigma^{ij}$ is the partonic level cross section for the collisions
$ij \rightarrow {\tilde q} {\tilde {\bar q}} ({\tilde g}{\tilde g})$ etc., where the indices
$i,j$ run over q, $\bar q$ and g. The partonic level center of mass energy ${\hat s}$ is related to the
hadronic level center of mass energy $s$ by: $\hat s~=~x_1x_2 s$ and
the partonic level squark and gluino production cross
sections $\sigma^{ij}(\hat s) $ can be obtained from the matrix element squares above
and are given by:

\begin{small}
\begin{eqnarray*}
  \sigma^B(q_i\qb_j\to\sq\sqb) & = &
  \delta_{ij} \,\frac{n_f\pi\as^2}{\hat s}\,\bs\left[\frac{4}{27}
  -\frac{16\ms^2}{27\hat s} \right] \\
  &&{} +\delta_{ij}\,\frac{\pi\as\hat{\alpha}_s}{\hat s}\left[\bs\left(\frac{4}{27}
  +\frac{8\md^2}{27\hat s}\right) +\left(\frac{8\mg^2}{27\hat s}
  +\frac{8\md^4}{27\hat s^2} \right)
   \Lg{\hat s+2\md^2-\hat s\bs}{\hat s+2\md^2+\hat s\bs} \right] \nonumber \\
  &&{}  +\frac{\pi\hat{\alpha}_s^2}{\hat s}\left[\bs\left(
  -\frac{4}{9}-\frac{4\md^4}{9(\mg^2\hat s+\md^4)}\right)
  +\left(-\frac{4}{9} -\frac{8\md^2}{9\hat s}\right)
   \Lg{\hat s+2\md^2-\hat s\bs}{\hat s+2\md^2+\hat s\bs} \right]\nonumber \\[0.2cm]
  \sigma^B(gg\to\sq\sqb) & = &
  \frac{n_f\pi\as^2}{\hat s}\left[
  \bs \left(\frac{5}{24} +\frac{31\ms^2}{12\hat s} \right)
  +\left(\frac{4\ms^2}{3\hat s}+ \frac{\ms^4}{3\hat s^2}\right)
  \log\left(\frac{1-\bs}{1+\bs}\right) \right]\\[0.2cm]
  \sigma^B(q_i q_j\to\sq\sq) & = &
  \frac{\pi\hat{\alpha}_s^2}{\hat s}\left[\bs\left(
  -\frac{4}{9}-\frac{4\md^4}{9(\mg^2\hat s+\md^4)}\right)
  +\left(-\frac{4}{9} -\frac{8\md^2}{9\hat s}\right)
   \Lg{\hat s+2\md^2-\hat s\bs}{\hat s+2\md^2+\hat s\bs} \right] \\
  & &{} + \delta_{ij}\,\frac{\pi\hat{\alpha}_s^2}{\hat s}
  \left[ \frac{8\mg^2}{27(\hat s+2\md^2)}
   \Lg{\hat s+2\md^2-\hat s\bs}{\hat s+2\md^2+\hat s\bs} \right] \nonumber\\[0.2cm]
  \sigma^B(q\qb\to\gl\gl) & = &
  \frac{\pi\as^2}{\hat s}\, \bg\left(\frac{8}{9} +\frac{16\mg^2}{9\hat s}\right)
  \\
  & &{} +\frac{\pi\as\hat{\alpha}_s}{\hat s}\left[
  \bg\left(-\frac{4}{3}-\frac{8\md^2}{3\hat s} \right)
  +\left(\frac{8\mg^2}{3\hat s} +\frac{8\md^4}{3\hat s^2}\right)
  \Lg{\hat s-2\md^2-\hat s\bg}{\hat s-2\md^2+\hat s\bg} \right] \nonumber\\
  & &{} +\frac{\pi\hat{\alpha}_s^2}{\hat s}\left[
  \bg\left(\frac{32}{27}+\frac{32\md^4}{27(\ms^2\hat s+\md^4)}\right)
  +\left(-\frac{64\md^2}{27\hat s} -\frac{8\mg^2}{27(\hat s-2\md^2)}\right)
   \Lg{\hat s-2\md^2-\hat s\bg}{\hat s-2\md^2+\hat s\bg} \right] \nonumber\\[0.2cm]
  \sigma^B(gg\to\gl\gl) & = &
  \frac{\pi\as^2}{\hat s}\left[
  \bg\left(-3-\frac{51\mg^2}{4\hat s}\right)
  + \left(-\frac{9}{4}
  -\frac{9\mg^2}{\hat s} +\frac{9\mg^4}{\hat s^2} \right)
  \log\left(\frac{1-\bg}{1+\bg} \right) \right] \\[0.2cm]
  \sigma^B(qg\to\sq\gl) & = &
  \frac{\pi\as\hat{\alpha}_s}{\hat s}\,\left[
  \frac{\kappa}{\hat s}\left(-\frac{7}{9} -\frac{32\md^2}{9\hat s}\right)
  + \left(-\frac{8\md^2}{9\hat s}+\frac{2\ms^2\md^2}{\hat s^2}
  +\frac{8\md^4}{9\hat s^2} \right)
  \Lg{\hat s-\md^2-\kappa}{\hat s-\md^2+\kappa} \qquad \right. \\
  & &\hphantom{\frac{\pi\as\hat{\alpha}_s}{\hat s}a} \left. {}
  + \left(-1-\frac{2\md^2}{\hat s}+\frac{2\ms^2\md^2}{\hat s^2}
  \right)  \Lg{\hat s+\md^2-\kappa}{\hat s+\md^2+\kappa} \right], \nonumber
\end{eqnarray*}
with
\begin{eqnarray}
  \bs  = \sqrt{1-\frac{4\ms^2}{\hat s}} \qquad
  \bg  = \sqrt{1-\frac{4\mg^2}{\hat s}} \\
  \md^2  = \mg^2 -\ms^2 \qquad
  \kappa  = \sqrt{(\hat s-\mg^2-\ms^2)^2-4\mg^2\ms^2} \\
  \as  = g_s^2/4\pi \qquad
  \hat{\alpha}_s  = \ghat^2/4\pi.
\end{eqnarray}
\end{small}

\begin{figure}[htb]
\vspace{2pt}
\centering{{\epsfig{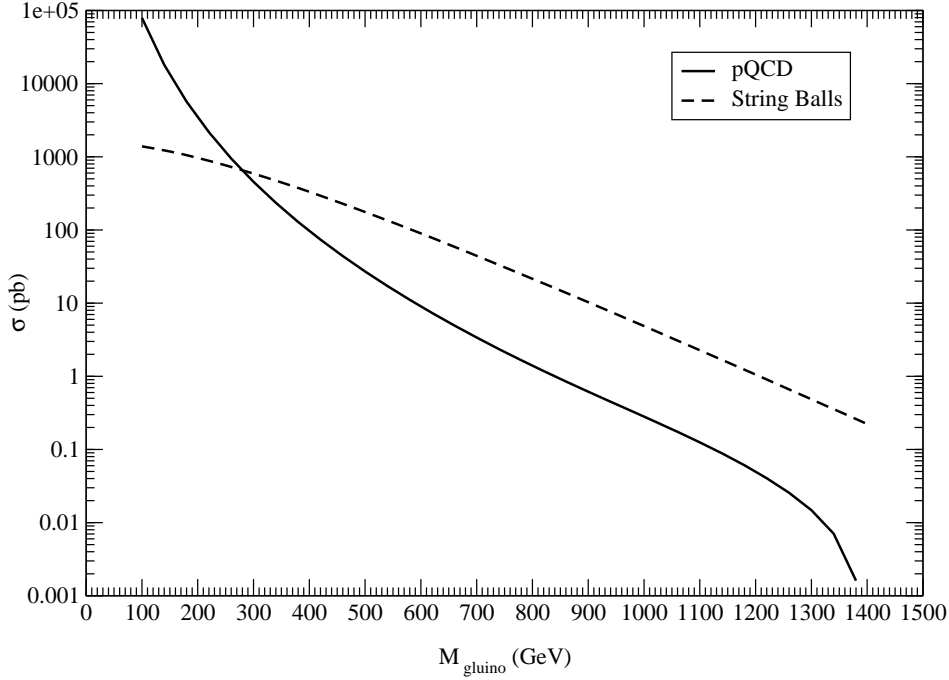}}}
\caption{ Total cross section for gluino production at LHC as a
function of gluino mass.}
\label{fig3}
\end{figure}

In our calculation we set the strong coupling equal to the QCD-SUSY coupling:
$g_s = \hat{g_s}$ and use the CTEQ6M PDF inside the proton \cite{cteq}.
We choose the factorization and renormalization scales to be $Q=m_{\tilde{q}}, m_{\tilde{g}}$,
(the squark and gluino masses respectively) and multiply a K factor of 1.5 to take into
account the higher order corrections.

\begin{figure}[htb]
\vspace{2pt}
\centering{{\epsfig{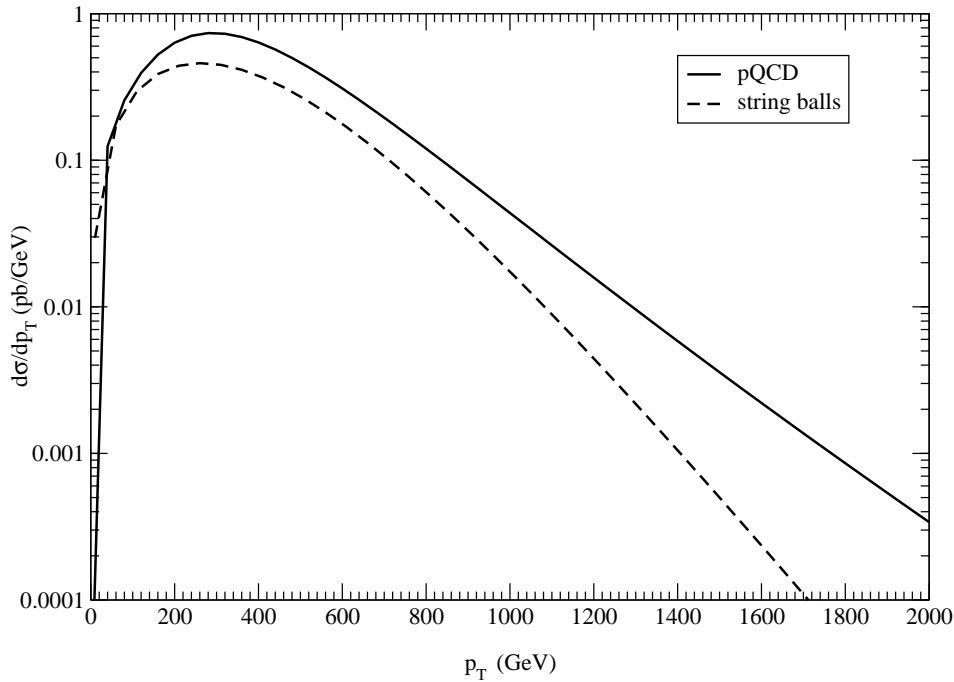}}}
\caption{ $p_T$ distribution of squark production at LHC.}
\label{fig4}
\end{figure}

\section{squark and gluino production From TeV Scale String Balls at LHC}

\begin{figure}[htb]
\vspace{2pt}
\centering{{\epsfig{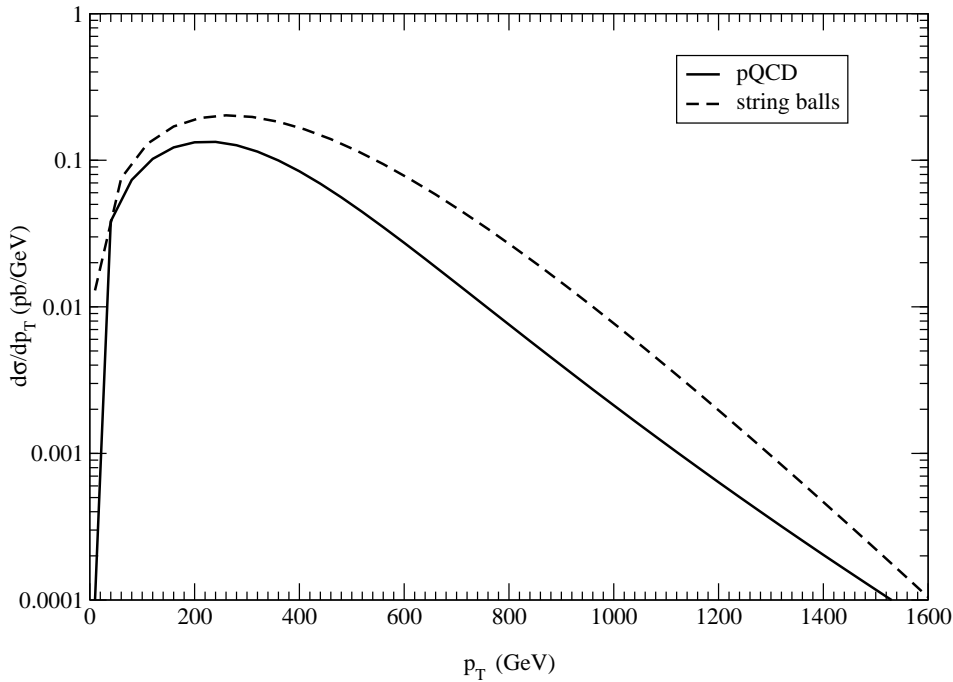}}}
\caption{ $p_T$ distribution of gluino production at LHC.}
\label{fig5}
\end{figure}

The differential cross section for squark production with mass $M_{\tilde q}$, momentum $\vec{p}$ and energy
$E =\sqrt{{\vec p}^2+M_{\tilde q}^2}$ from string ball of temperature $T_{SB}$ at LHC is given by \cite{arif,nayaks}
\bea
&& \frac{Ed\sigma_{\rm squark}}{d^3p}
= \frac{1}{(2\pi)^3s}{\sum}_{ab}~\int_{M^2_{s}}^{\frac{M^2_s}{g^4_s}}~dM^2~
\int \frac{dx_a}{x_a} ~f_{a/p}(x_a, \mu^2)~f_{b/p}(\frac{M^2}{sx_a}, \mu^2)~\hat{\sigma}^{ab}(M)~\frac{A_n c_n \gamma \tau_{SB} p^\mu u_\mu}{(e^{\frac{p^\mu u_\mu}{T_{SB}}} - 1)}\,, \nonumber \\
\label{bksh}
\eea
where $M_s$ is the string mass scale and $g_s$ is the string coupling which is less than $1$ for the string perturbation theory to be valid,
see eq. (\ref{scale}). The flow velocity is $u^\mu$ and $\gamma$ is the Lorentz boost factor with
\bea
\gamma {\vec v}_{SB} = (0,0,\frac{(x_1-x_2) \sqrt{s}}{2M_{SB}}).
\label{4v}
\eea
$A_n$ is the $d(=n+3)$ dimensional area factor \cite{bv,nayaks}. We will use the number of extra dimensions
$n=6$ in our calculation. The partonic level string ball production cross section $\hat{\sigma}^{ab}$
is given by eq. (\ref{sbc}). We have used CTEQ6M PDF \cite{cteq} in our calculation.

Similarly, the differential cross section for gluino production with mass $M_{\tilde g}$, momentum $\vec{p}$ and energy
$E =\sqrt{{\vec p}^2+M_{\tilde g}^2}$ from string ball of temperature $T_{SB}$ at LHC is given by
\bea
&& \frac{Ed\sigma_{\rm gluino}}{d^3p}
= \frac{1}{(2\pi)^3s}{\sum}_{ab}~\int_{M^2_{s}}^{\frac{M^2_s}{g^4_s}}~dM^2~
\int \frac{dx_a}{x_a} ~f_{a/p}(x_a, \mu^2)~f_{b/p}(\frac{M^2}{sx_a}, \mu^2)~\hat{\sigma}^{ab}(M)~\frac{A_n c_n \gamma \tau_{SB} p^\mu u_\mu}{(e^{\frac{p^\mu u_\mu}{T_{SB}}} + 1)}\,. \nonumber \\
\label{gbksh}
\eea

\section{Results and Discussions}

In this section we present the results of our calculation in $pp$ collisions
at LHC at $\sqrt s$ = 14 TeV.

\begin{figure}[htb]
\vspace{2pt}
\centering{{\epsfig{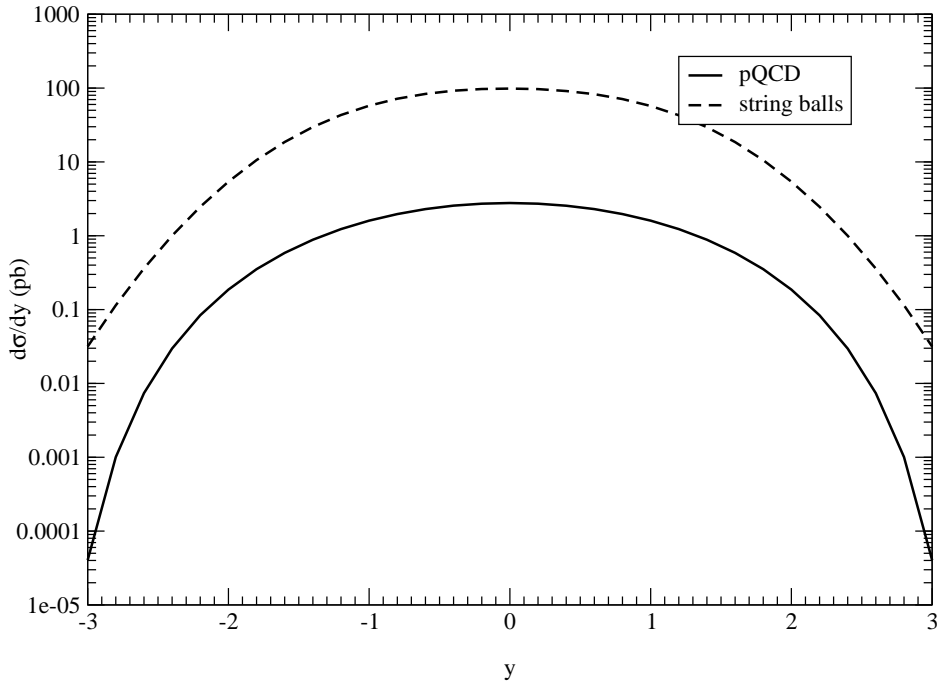}}}
\caption{ Rapidity distribution of squark production at LHC.}
\label{fig6}
\end{figure}

In Fig. 2 we contrast the result for squark
production from the pQCD-SUSY calculation with that from thermal
string ball emission as a
function of squark mass in pp collisions at $\sqrt s$ = 14 TeV at LHC.
The solid line is squark production from the pQCD-SUSY calculation with
$m_{\tilde{q}}/m_{\tilde{g}}$=0.8.
The dashed line is squark production from string balls at
with string mass scale $M_s$= 1 TeV. It can be seen that if the string mass
scale $M_s\sim$ 1 TeV then the squark production from string balls at LHC
is much larger than that from pQCD-SUSY processes for squark mass larger than
200 GeV.

\begin{figure}[htb]
\vspace{2pt}
\centering{{\epsfig{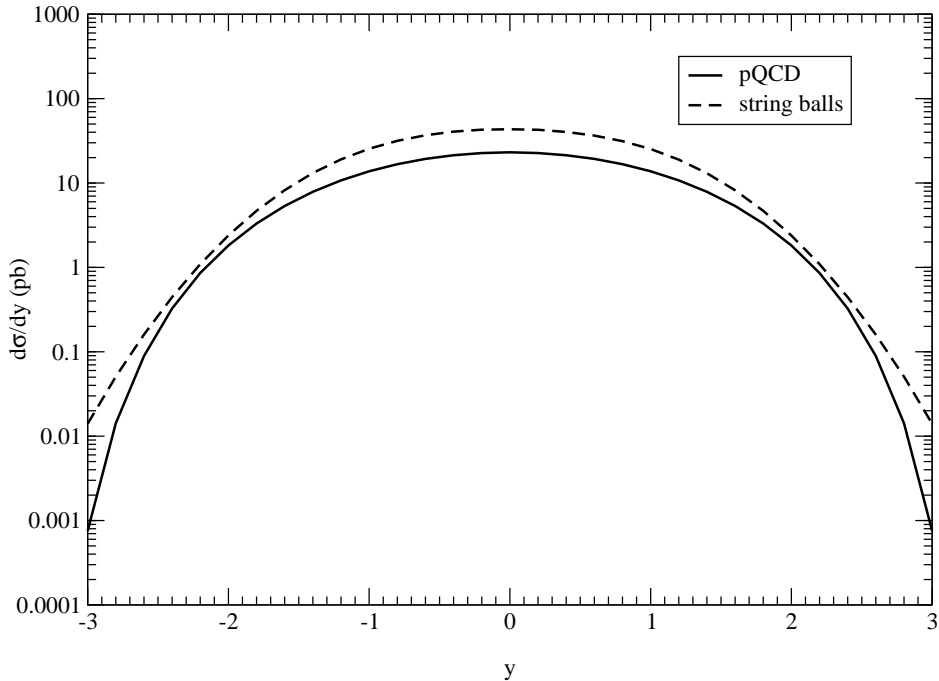}}}
\caption{ Rapidity distribution of gluino production at LHC.}
\label{fig7}
\end{figure}

In Fig.3 we contrast the result for gluino
production from the pQCD-SUSY calculation with that from thermal
string ball emission as a
function of gluino mass in pp collisions at $\sqrt s$ = 14 TeV at LHC.
The solid line is gluino production from the pQCD-SUSY calculation with
$m_{\tilde{q}}/m_{\tilde{g}}$=0.8.
The dashed line is gluino production from string balls
with string mass scale $M_s$= 1 TeV. It can be seen that if the string mass
scale $M_s\sim$ 1 TeV then the gluino production from string balls at LHC
is much larger than that from pQCD-SUSY processes for gluino mass larger than
300 GeV.

In Fig.4 we present $\frac{d\sigma}{dp_T}$ for the squark production cross section,
both from pQCD-SUSY processes and from string balls at LHC. The solid line is the squark
production cross section from direct pQCD-SUSY
production processes as a function of $p_T$ at LHC at $\sqrt s$ = 14 TeV pp collisions.
Here we have taken the squark mass to be 500 GeV with $m_{\tilde{q}}/m_{\tilde{g}} = 0.8$.
The dashed line is the $p_T$ distribution of squark production from string balls
with string mass scale $M_s$= 1 TeV.

In Fig.5 we present $\frac{d\sigma}{dp_T}$ for the gluino production cross section,
both from pQCD-SUSY processes and from string balls at LHC. The solid line is the gluino
production cross section from direct pQCD-SUSY
production processes as a function of $p_T$ at LHC at $\sqrt s$ = 14 TeV pp collisions.
Here we have taken the gluino mass to be 500 GeV with $m_{\tilde{q}}/m_{\tilde{g}} = 0.8$.
The dashed line is the $p_T$ distribution of gluino production from string balls
with string mass scale $M_s$= 1 TeV.

In Fig.6 we present the rapidity distribution results for squark
production both from pQCD-SUSY processes and from string balls at LHC.
The solid line is the squark production cross section from direct pQCD-SUSY
production processes as a function of rapidity at LHC at $\sqrt s$ = 14 TeV pp collisions.
Here we have taken squark mass to be 500 GeV with $m_{\tilde{q}}/m_{\tilde{g}} = 0.8$.
The rapidity range covered is from -3 to 3.
The dashed line is the rapidity distribution of squark production from string balls
at LHC with string mass scale $M_s$= 1 TeV.

In Fig.7 we present the rapidity distribution results for gluino
production both from pQCD-SUSY processes and from string balls at LHC. The solid line is the gluino
production cross section from direct pQCD-SUSY
production processes as a function of rapidity at LHC at $\sqrt s$ = 14 TeV pp collisions.
Here we have taken $m_{\tilde{q}}/m_{\tilde{g}} = 0.8$.
The rapidity range covered is from -3 to 3. We have chosen the gluino mass to be 500 GeV.
The dashed line is the rapidity distribution of gluino production from string balls
at LHC with string mass scale $M_s$= 1 TeV.

From the above results it can be seen that there are significant squark and gluino production
from string balls at LHC. Hence, in the absence of black hole production at LHC, an enhancement
in the supersymmetry production may be a signature of TeV scale string physics at LHC.

\section{Conclusions}

If extra dimensions are found in the second run of LHC in the $pp$ collisions at
$\sqrt{s}$ = 14 TeV then the string scale can be $\sim$ TeV, and we should produce
string balls at LHC. In this paper we have studied supersymmetry (squark and gluino)
production from string balls at LHC in $pp$ collisions at $\sqrt{s}$ = 14 TeV
and have compared that with the parton fusion results using pQCD. We have found
significant squark and gluino production from string balls at LHC which is
comparable to parton fusion pQCD results. Hence, in the absence of black
hole production at LHC, an enhancement in supersymmetry production
can be a signature of TeV scale string physics at LHC.

The LHC will also collide two lead nuclei at $\sqrt{s}$ = 5.5 TeV per nucleon
to produce quark-gluon plasma \cite{qgp,qgp1,qgp2,qgp3}. Since the total energy in the two lead nuclei collisions
at LHC will be $\sim$1150 TeV, we may expect to see new physics \cite{newph} in the nuclear collisions
at LHC as well.

\end{document}